\documentclass{iauc}
\usepackage{graphicx}

\title[SDI pipeline] 
{Suppressing Speckle Noise for Simultaneous Differential Extrasolar Planet 
Imaging (SDI) at the VLT and MMT}

\author[Biller et al.]   
{Beth A. Biller$^1$, Laird M. Close$^1$, 
Rainer Lenzen$^{2}$, Wolfgang Brandner$^{2}$, Donald McCarthy$^1$, 
Eric Nielsen$^1$, Stephan Kellner$^{2}$, and Markus Hartung$^3$}

\affiliation{$^1$Steward Observatory, University of Arizona, Tucson, AZ 85721
\break email: bbiller@as.arizona.edu\\[\affilskip]
$^2$ Max-Planck-Institut f\"ur Astronomie, K\"onigstuhl 17, 69117 
Heidelberg, Germany \break
$^3$ European Southern Observatory, Alonso de Cordova 3107, Santiago 19, Chile
\break}

\pubyear{2005}
\volume{200} 
\pagerange{1--5}
\date{?? and in revised form ??}
\jname{Direct Imaging of Exoplanets: Science and Techniques}
\editors{C. Aime and F. Vakili, Eds.}
\begin{document}

\maketitle

\begin{abstract}

We discuss the instrumental and data reduction techniques used to suppress
speckle noise with the Simultaneous Differential Imager (SDI)
implemented at the VLT and the MMT.  SDI uses a double Wollaston prism and 
a quad filter to take 4 identical images simultaneously
at 3 wavelengths surrounding the 1.62 $\mu$m methane bandhead found in 
the spectrum of cool brown dwarfs and gas giants.  By performing 
a difference of images in these filters, speckle noise from
the primary can be significantly attenuated, resulting in photon noise limited 
data past 0.5''.  Non-trivial data reduction tools 
are necessary to pipeline the simultaneous differential imaging.
Here we discuss a custom algorithm implemented in IDL to 
perform this reduction.
The script performs basic data reduction tasks but also precisely aligns
images taken in each of the filters using a custom shift and subtract routine. 
In our survey of nearby young stars at the VLT and MMT (see
Biller et al., this conference), we achieved H band contrasts
$>$25000 (5$\sigma$ $\Delta$F1(1.575 $\mu$m) $>$ 10.0 mag, 
$\Delta$H$>$11.5 mag for a T6 spectral type object) 
at a separation of 0.5" from the primary star.  
We believe that our SDI images are among the highest contrast astronomical 
images ever made from ground or space for methane rich companions.

\keywords{instrumentation: adaptive optics, methods: data analysis, 
techniques: image processing, (stars:) planetary systems}
\end{abstract}

\firstsection 
\section{Introduction}

In theory, adaptive optics (AO) 
systems that are ``photon noise limited'' can detect an object up to 
10$^{5-6}$ times fainter its primary at separations of $\sim$1''
 -- sufficient to detect a young, 
self-luminous giant extrasolar planet.  However, overcoming the large
contrast difference between star and planet
is not the only obstacle in directly detecting extrasolar planets -- 
all AO also systems suffer from a 
limiting ``speckle noise'' floor (Racine et al. 1999).  Within 1'' 
of the primary star, the field is filled with speckles left over from 
instrumental features and residual atmospheric turbulence after adaptive 
optics correction.  These speckles vary as a function of time and color.  
For photon noise limited data, the signal to noise S/N increases as 
t$^{0.5}$, 
where t is the exposure time.  For speckle-noise limited data, the S/N 
does 
not increase with time past a specific speckle-noise floor (limiting 
contrasts to $\sim$10$^3$ at 0.5'').  This speckle-noise floor is 
considerably above the photon noise limit and makes planet detection very 
difficult.  Interestingly, space telescopes such as HST also suffer from a 
somewhat similar limiting speckle-noise floor due to imperfect optics and 
``breathing'' (Schneider et al. 2003).  Direct detection of 
extrasolar giant planets requires special instrumentation to suppress 
this speckle noise floor and produce photon noise limited images. 
 
Simultaneous Differential Imaging is an instrumental method which can be 
used to calibrate and remove the ``speckle noise'' in AO images, while 
also isolating the planetary light from the starlight.  This method was 
pioneered by Racine et al. (1999), Marois et al. (2000), Marois et al. 
(2002) and Marois et al. (2005).  
It exploits the fact that all cool (T$_{eff}$ $<$1200 K) 
extra-solar giant planets are thought to 
have strong CH$_4$ (methane) absorption 
redwards of 1.62 $\mu$m in the H band infrared atmospheric window 
(Burrows et al. 2001, Burrows et al. 2003).  Our SDI device obtains four 
images of a star simultaneously through three slightly different narrowband
 filters (sampling both inside and outside of the CH$_4$ features).  These 
images are then differenced.  This subtracts out the halo and speckles from 
the bright star to reveal any massive extrasolar planets orbiting that star.  
Since a massive planetary companion will be brightest in the F1(1.575 $\mu$m)
filter and absorbed in the rest, while the star is bright in all three, 
a difference can be chosen which subtracts out the star's light 
and reveals the light from the companion.  Thus, SDI also helps 
eliminate the large contrast difference between the star and substellar 
companions. 

  \begin{figure}
   \begin{center}
   \begin{tabular}{cc}
   \includegraphics[height=3cm]{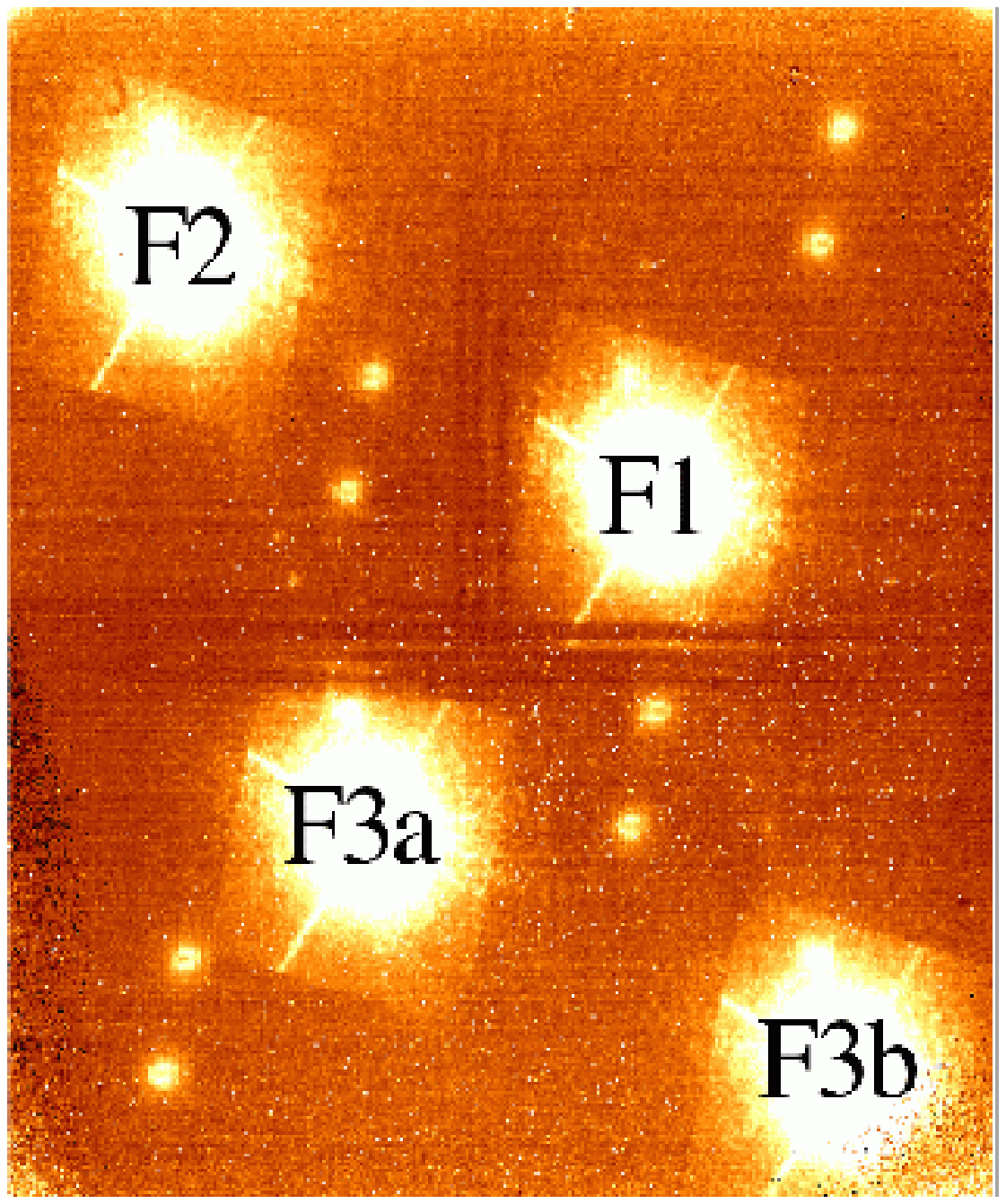} &
   \includegraphics[height=3cm]{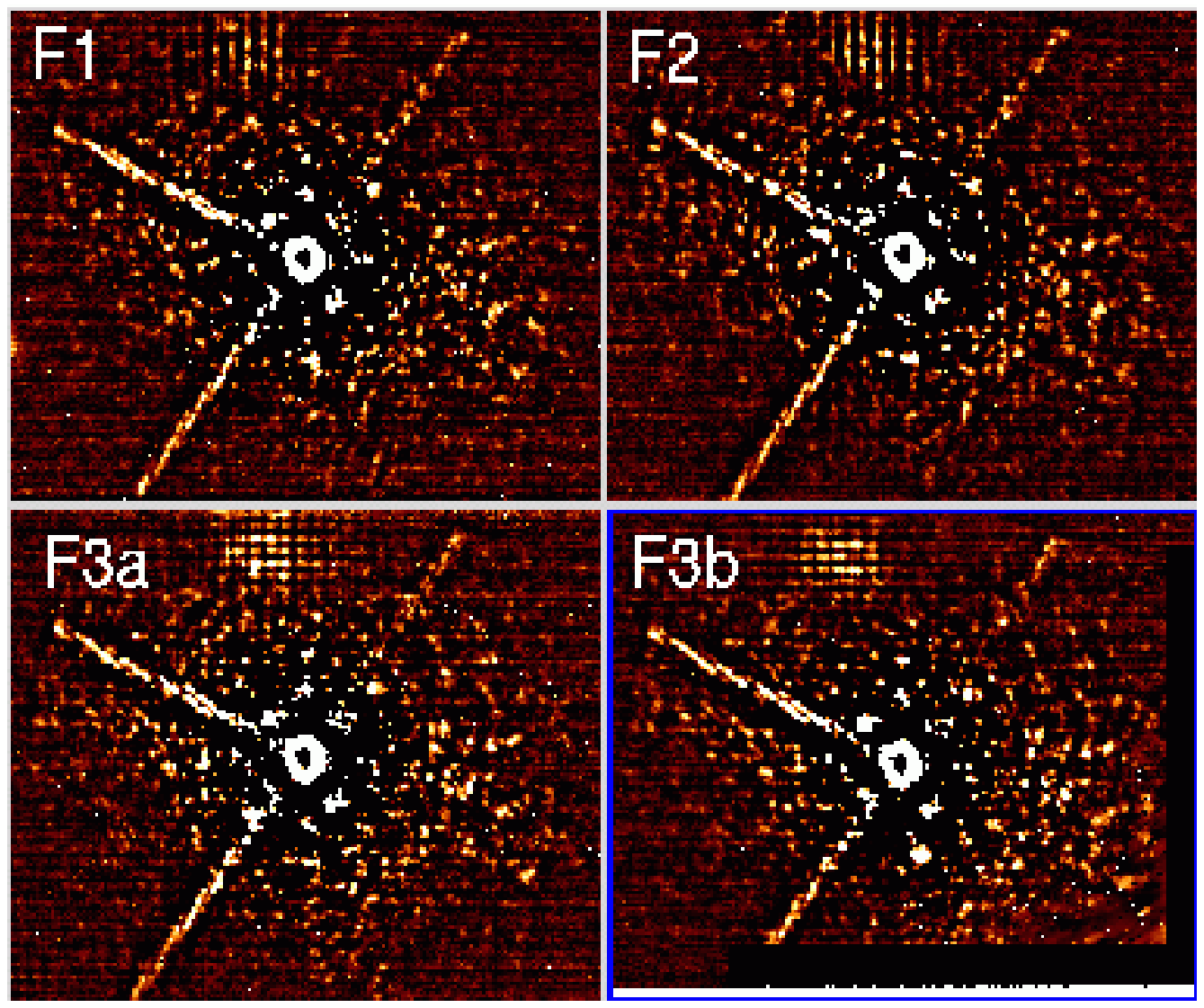} \\
   \end{tabular}
   \end{center}
   \caption[Raw VLT SDI data]
   { \label{fig:SDIRAW} Left: Two minutes of raw SDI data from NACO SDI's 
1024$\times$1024 Aladdin array 
in the CONICA AO camera (Lenzen et al. 2004).  Right: 
Same dataset, slightly processed.  Apertures have been selected around 
each filter image.  In
order to reveal the speckle pattern, a heavily smoothed image was subtracted
from the raw images (unsharp masking).  
The resulting speckle patterns are very similar between the 4 simultaneous 
images which means that an effective subtraction of 
speckles can be obtained between the filters.
}
   \end{figure}

\section{The SDI AO Cameras at the VLT and MMT}
SDI optics are currently implemented at the 6.5m MMT (using the 
MMT AO adaptive secondary mirror and the ARIES AO 
camera -- McCarthy et al. 1998) and at the ESO VLT (using the 8m UT4 and the 
NAOS-CONICA (NACO) AO system) by a group headed by L. Close and R. 
Lenzen (Close et al. 2005, Lenzen et al. 2004, Lenzen et al. 2005).  
Both devices are fully commissioned and available for observing.

The SDI technique requires some specialized optics consisting of a cryogenic
custom 
double calcite Wollaston device and a focal plane quad CH$_4$ filter. Our 
custom Wollaston splits the beam into four identical beams while minimizing 
non-common path errors.  The {\it differential} non-common path errors
are less than 20\,nm~RMS per Zernike mode between the beams 
(Lenzen et al. 2004).
Each of the four beams is fed through one of 
the filters on the quad filter.  Filter wavelengths were chosen on and off
the methane absorption feature at 1.62 $\mu$m and were spaced closely 
(every 0.025 $\mu$m) in order to limit residuals due to speckle and 
calcite chromatism.
We used four filters F1, F2, F3a, and F3b with central wavelengths 
F1=1.575 $\mu$m, F2=1.600 $\mu$m, and F3a=F3b=1.625 $\mu$m.  
The filters are approximately 0.025 $\mu$m in bandwidth.   
A cold 5''$\times$5'' focal plane mask has been implemented as a field stop 
for the VLT device.  No coronagraph is currently used, since the Strehl 
ratios ($\sim$20-30$\%$) are too low to increase the contrasts 
significantly.  The special f/40 SDI camera has a platescale of 0.017$''$/pix
at the VLT and 0.02$''$/pix at the MMT.

The SDI device has 
already produced a number of important scientific results: the discovery 
of AB Dor C (Close et al. 2005) which is the tightest (0.16'') 
low mass companion detected by direct imaging, detailed 
surface maps of Titan 
(Hartung et al. 2004), the discovery of $\epsilon$ Indi Ba-Bb, the 
nearest binary brown dwarf (McCaughrean et al. 2005), and evidence 
of orbital motion for Gl 86B, the first known 
white dwarf companion to an exoplanet host star
(Mugrauer and Neuh\"auser 2005).

\section{Observational Technique and Data Reductions}

A raw dataset from NACO SDI is shown in Fig.~\ref{fig:SDIRAW}.  
The inner 0.2'' diameter core is saturated in each image to increase signal 
in the halo.  After unsharp masking, we find that the speckle patterns 
in each of the separate filters are nearly identical.  
(See Fig.~\ref{fig:SDIRAW}). 
This bodes well for our ability to attenuate speckle noise. 
 
\begin{figure}
   \begin{center}
   \begin{tabular}{cc}
   \includegraphics[height=9cm]{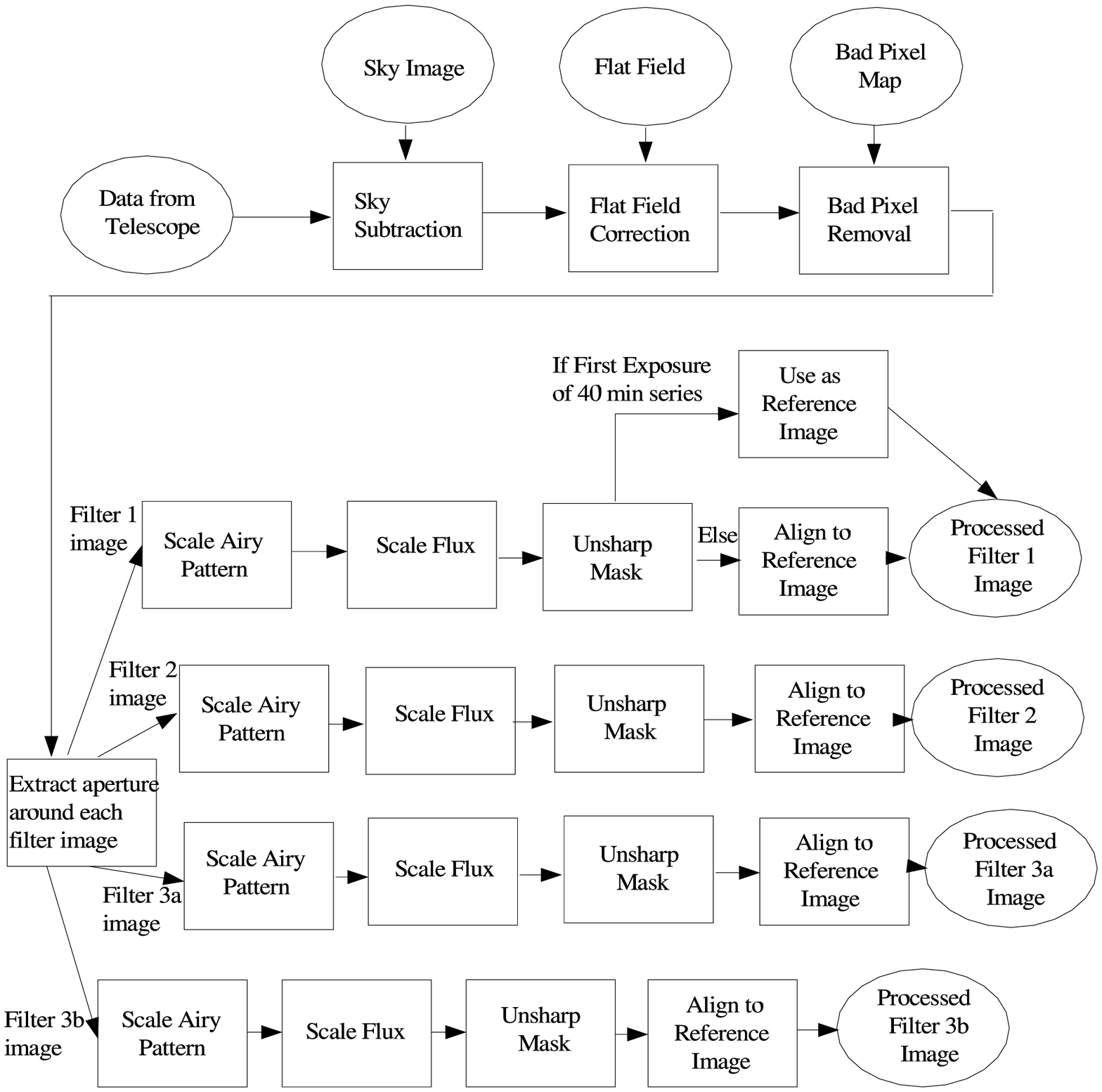} &
   \includegraphics[height=9cm]{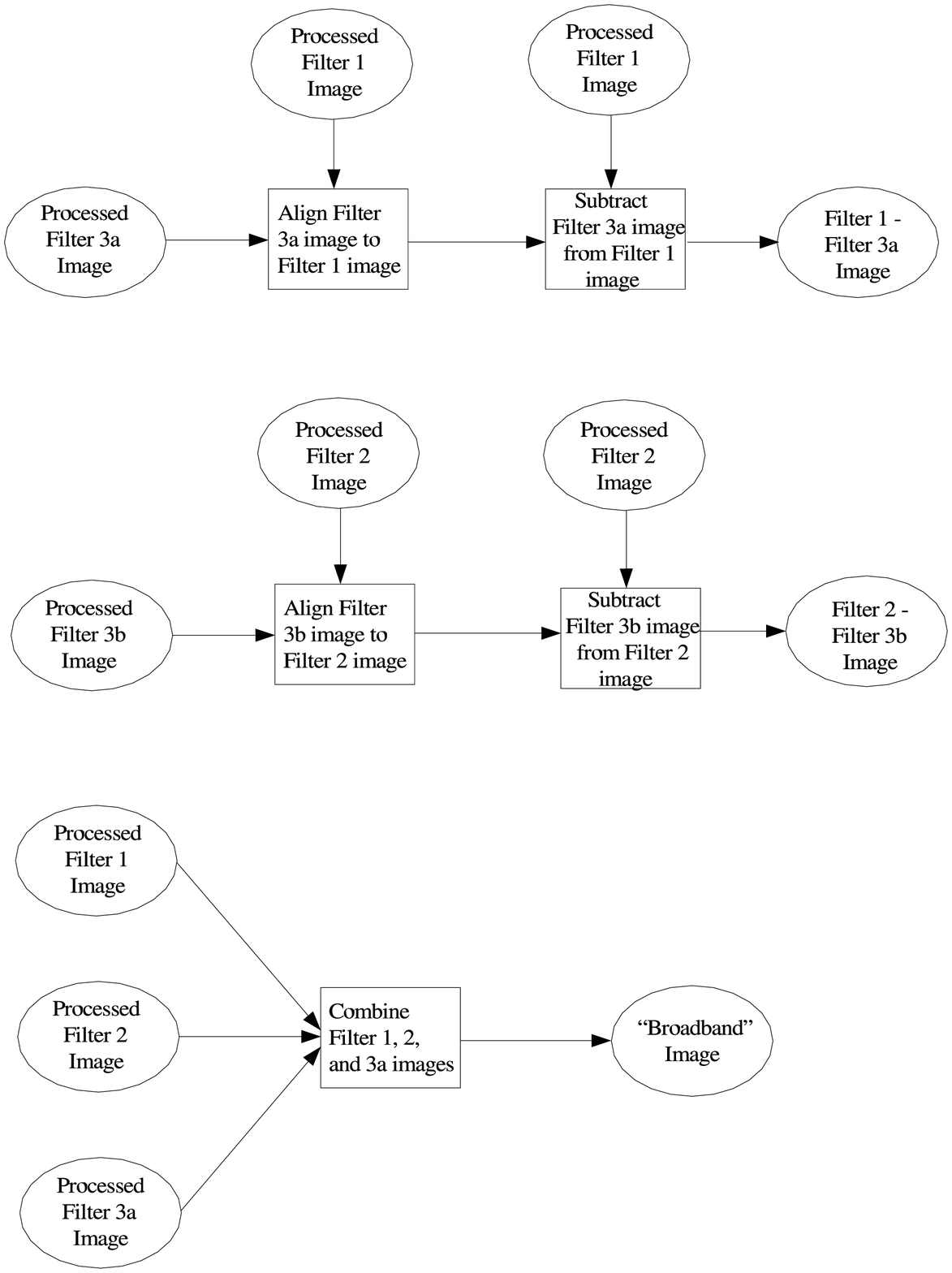} \\
   \end{tabular}
   \end{center}
   \caption[Pipeline Block Diagram]
   { \label{fig:pipeline} 
	Pipeline Block Diagrams
}
   \end{figure}

To distinguish between faint planets and any residual speckles, we observe 
each object at a variety of position angles (usually 
a series of 0$^{\circ}$ and 33$^{\circ}$ observations).  Instrumental and 
telescope 
``super speckles'' (Racine et al. 1999) should not rotate with a change of 
rotator angle; however, a real planet should appear to rotate by the change in 
rotator angle.  The data is reduced using a custom IDL script. 
A pipeline block diagram for this IDL script is presented in 
Fig.~\ref{fig:pipeline}.  Alignments are performed using a custom 
shift and subtract algorithm.  We calculate 2 
differences (and one non-differenced combination) which are sensitive to 
substellar companions of spectral types ``T'' 
(T$_{eff}$ $<$ 1200 K), ``Y'' (T$_{eff}$ $\leq$ 600 K), 
and ``L'' (T$_{eff}$ $>$ 1200 K). 
Data taken at different position angles are subtracted (e.g. 20 minutes of 
data at 0 degrees minus 20 minutes of data at 33 degrees) in order to further
attenuate speckle noise.    

\begin{figure}
   \begin{center}
   \begin{tabular}{ccc}
   \includegraphics[height=3.8cm]{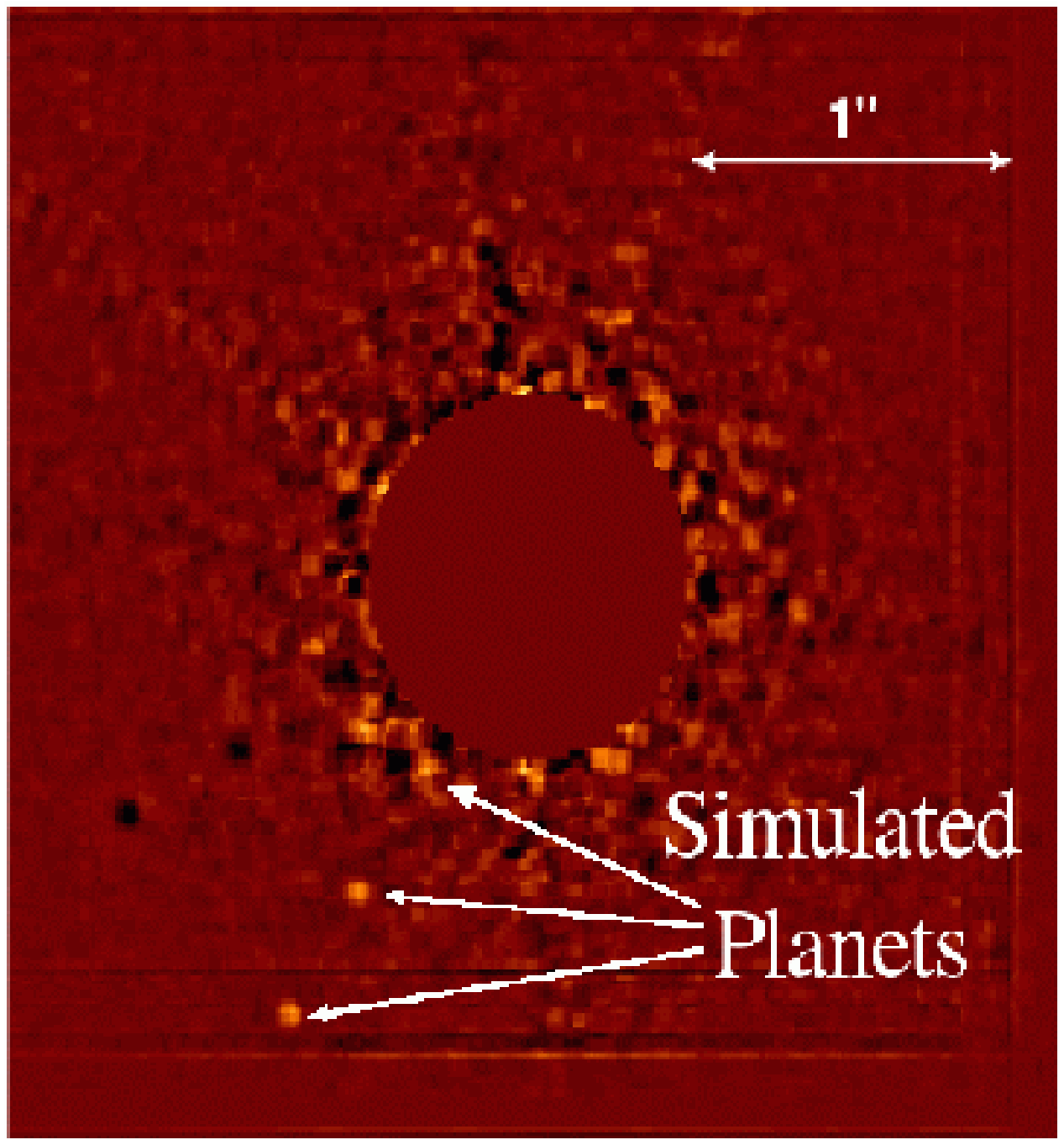} &
   \includegraphics[height=3.8cm]{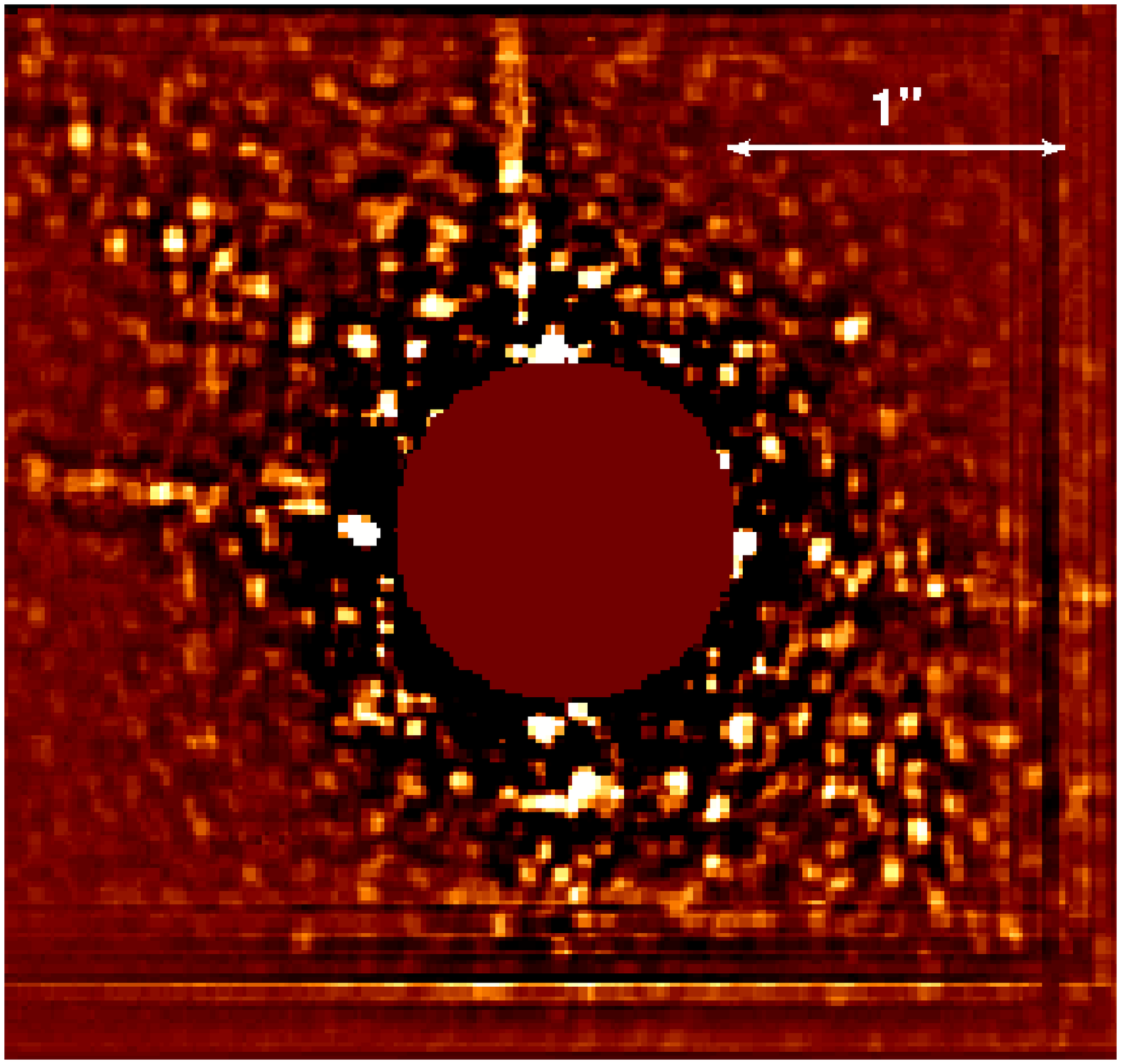} &
   \includegraphics[height=4cm]{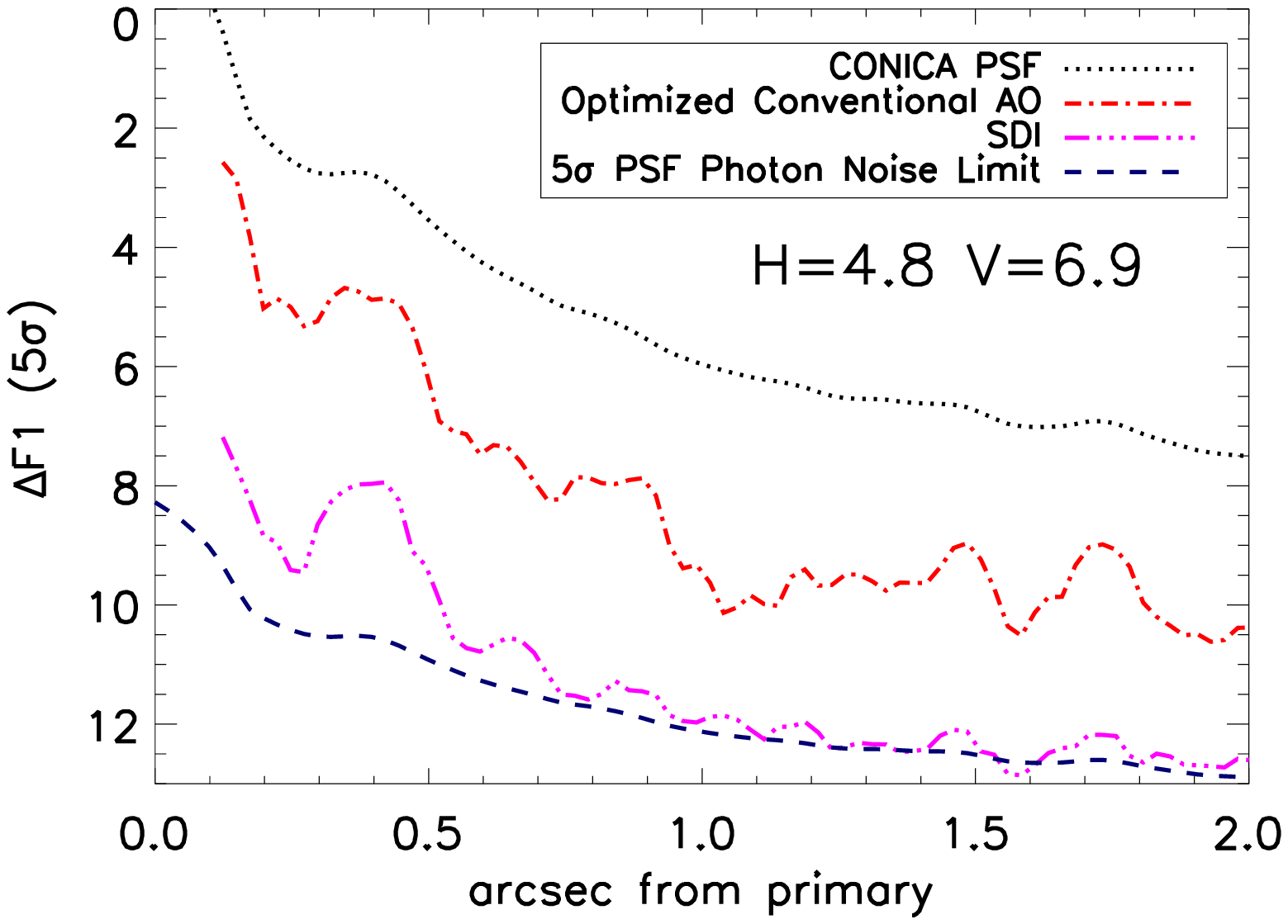} \\
   \end{tabular}
   \end{center}
   \caption[Reduced VLT SDI data]
   { \label{fig:SDIRED} {\bf Left:} A complete reduced dataset 
(40 minutes of data at a series of rotator angles -- 
0$^{\circ}$, 33$^{\circ}$, 33$^{\circ}$, 0$^{\circ}$) from the VLT SDI device.
Simulated planets have been added at separations of 
0.55, 0.85, and 1.35'' from the primary, with $\Delta$F1 = 10 mag 
(attenuation in magnitudes in the F1 1.575 $\mu$m 
filter) fainter than the primary.
  These planets are scaled from unsaturated images of the example star 
taken right before the example dataset (and have fluxes in each filter 
appropriate for a T6 object).  Past 0.7'', the simulated planets are 
detected with S/N $>$ 10.
{\bf Center:} Standard AO data reduction of the same
dataset.  Images have been coadded, flat-fielded, sky-subtracted, and 
unsharp-masked.  Simulated planets have been added with the 
same properties and at the same separations as before.  None of the simulated
planets are clearly 
detected in the standard AO reduction.  Additionally, numerous
bright super speckles remain in the field.
{\bf Right:} $\Delta$F1 (5$\sigma$ noise level in the 1.575 $\mu$m
F1 filter) vs. Separation for 40 minutes of VLT SDI 
data for a 70 Myr K1V star at 15 pc. The top curve is the AO PSF.  The next 
curve is the ``classical AO PSF'' unsharp masked.  The third curve down is 
40 minutes of SDI 0$^{\circ}$-33$^{\circ}$ data.
The last curve is the theoretical contrast
limit due to photon-noise.  At star-companion separations $>$ 0.5'', 
we are photon-noise limited and achieve star to planet 
H band contrasts $>$25000 (5$\sigma$ $\Delta$F1(1.575 $\mu$m) $>$ 10.0 mag, 
$\Delta$H$>$11.5 mag for a T6 spectral type object) 
at a separation of 0.5" from the primary star.  
}
   \end{figure}

A fully reduced dataset from the VLT SDI device as well as the 
same dataset reduced in a standard AO manner is presented in 
Fig.~\ref{fig:SDIRED}.  Simulated planets were inserted into the dataset
pre-reduction.
In the SDI reduction, simulated planets with $\Delta$F1=10  
(attenuation in magnitudes in the 1.575 $\mu$m
F1 filter) are detected with S/N $>$ 10 past 0.7''.  In comparison,
none of the simulated planets 
are clearly detected in the standard AO data reduction and numerous 
bright super speckles remain in the field.  A plot of $\Delta$F1 (for 
5$\sigma$ detections) 
vs. separation from the primary is presented in 
Fig.~\ref{fig:SDIRED}.
For this dataset, we achieved star to planet H band contrasts (5$\sigma$) 
$>$25000 (5$\sigma$ $\Delta$F1(1.575 $\mu$m) $>$ 10.0 mag, 
or $\Delta$H$>$11.5 mag for a T6 spectral type object) 
at a separation of 0.5" from the primary star -- approaching 
the photon-noise limit in 40 minutes of data.   
$\Delta$F1(1.575 $\mu$m) and $\Delta$H (for a methane object) 
for 3 of our survey stars (see Biller et al. this conference) as well 
as for two other comparison objects are shown in Table~\ref{tab:properties} -- 
it is clear that the achievable contrast varies according
to the magnitude of the object and total exposure time.  
We believe these are among  
the most sensitive astronomical images taken to date 
for methane rich companions.

\begin{table}\def~{\hphantom{0}}
  \begin{center}
  \caption{Properties of Example SDI Survey Stars and Comparison Stars}
  \label{tab:properties}
  \begin{tabular}{lcccccccc}\hline
      Case  & Spectral Type  &  Age & Distance & H & V & Exposure Time & $\Delta$F1$^{1}$
& $\Delta$H$^{1}$ \\\hline
	A   &  K2V           &  30 Myr & 45.5 pc & 7.1 & 9.1 & 40 min & 10.5 & 12 \\
	B   &  K1V	     &  70 Myr & 15 pc & 4.8 & 6.9 & 40 min & 10.5 & 12 \\
	C   &  M3V	     &  30 Myr & 24 pc & 7.1 & 12.2 & 40 min & 10 & 11.5 \\ 
10 late K-M stars$^{2}$ & K-M & 0-1 Gyr & 10-50 pc & 6.4-8.7 & 8-12 & 10-25 min & 8.61 & 10.31 \\ 
  Gl 86$^{3}$     &  K1V &  10 Gyr & 10.9 pc & 4.2 & 6.2 & 80 min & 12.8 & 14.3 \\ \hline   
  \end{tabular}
  \end{center}
~$^1$ 5$\sigma$ at 0.5'' ~$^2$ Masciadri et al. IAUC 200 ~$^3$ Mugrauer \& Neuh\"auser 2005
\end{table}

\begin{acknowledgments}
BAB acknowledges support through the NASA GSRP program.  LMC acknowledges 
support through NSF CAREER and NASA Origins grants.
\end{acknowledgments}

\end{document}